\documentclass[aps,twocolumn,showpacs,superscriptaddress,floatfix]{revtex4}
\usepackage{graphicx}
\usepackage{psfrag}
\usepackage{color}

\usepackage{amsmath}
\usepackage{amssymb}   
\usepackage{amstext}
\usepackage{amsthm}
\usepackage{amsfonts}
\usepackage{amsxtra}

\begin{document}

\newcommand{\ket}[1]{\left\vert #1\,\right\rangle}
\newcommand{\ketrc}{\left\vert\textsc{rc}\,\right\rangle}
\newcommand{\bra}[1]{\left\langle #1\,\right\vert}
\newcommand{\bbra}[1]{\left\langle\left\langle #1\,\right\vert\right.}
\newcommand{\brarc}{\left\langle\,\textsc{rc}\right\vert}
\newcommand{\braket}[2]{\left\langle #1\,\vert #2\,\right\rangle}
\newcommand{\bbraket}[2]{\left\langle\left\langle #1\,\vert #2\,\right\rangle\right.}
\newcommand{\de}{\partial}
\newcommand{\eps}{\varepsilon}
\newcommand{\tr}{\operatorname{\mathrm{Tr}}}
\newcommand{\re}{\mathrm{Re}}
\newcommand{\im}{\mathrm{Im}}
\newcommand{\R}{\mathbb{R}}
\newcommand{\eav}[1]{\left\langle #1\right\rangle}
\newcommand{\beq}{\begin{equation}}
\newcommand{\eeq}{\end{equation}}
\newcommand{\ben}{\begin{eqnarray}}
\newcommand{\een}{\end{eqnarray}}
\newcommand{\bea}{\begin{array}}
\newcommand{\eea}{\end{array}}
\newcommand{\om}{(\omega )}
\newcommand{\bef}{\begin{figure}}
\newcommand{\eef}{\end{figure}}
\newcommand{\leg}[1]{\caption{\protect\rm{\protect\footnotesize{#1}}}}
\newcommand{\ew}[1]{\langle{#1}\rangle}
\newcommand{\be}[1]{\mid\!{#1}\!\mid}
\newcommand{\no}{\nonumber}
\newcommand{\etal}{{\em et~al }}
\newcommand{\geff}{g_{\mbox{\it{\scriptsize{eff}}}}}
\newcommand{\da}[1]{{#1}^\dagger}
\newcommand{\cf}{{\it cf.\/}\ }
\newcommand{\ie}{{\it i.e.\/}\ }   
\setlength\abovedisplayskip{5pt}
\setlength\belowdisplayskip{5pt}

\title{Cooperative Robustness to Static Disorder: Superradiance and
  localization in a nanoscale ring to model natural light-harvesting systems}

\author{G.~Luca \surname{Celardo}}
\affiliation{Dipartimento di Matematica e
Fisica and Interdisciplinary Laboratories for Advanced Materials Physics,
 Universit\`a Cattolica del Sacro Cuore, via Musei 41, I-25121 Brescia, Italy}
\affiliation{ Istituto Nazionale di Fisica Nucleare,  Sezione di Pavia, 
via Bassi 6, I-27100,  Pavia, Italy}
\author{Giulio~G. \surname{Giusteri}}
\affiliation{Dipartimento di Matematica e
Fisica and Interdisciplinary Laboratories for Advanced Materials Physics,
 Universit\`a Cattolica del Sacro Cuore, via Musei 41, I-25121 Brescia, Italy}
\affiliation{International Research Center on Mathematics \& Mechanics of Complex Systems, via XIX marzo 1, I-04012 Cisterna di Latina, Italy}
\author{Fausto \surname{Borgonovi}}
\affiliation{Dipartimento di Matematica e
Fisica and Interdisciplinary Laboratories for Advanced Materials Physics,
 Universit\`a Cattolica del Sacro Cuore, via Musei 41, I-25121 Brescia, Italy}
\affiliation{ Istituto Nazionale di Fisica Nucleare,  Sezione di Pavia, 
via Bassi 6, I-27100,  Pavia, Italy}

\begin{abstract}                
We analyze a 1-d ring structure composed of many
two-level systems, in the limit where only one excitation is present.
The two-level systems are coupled to a common environment, where the
excitation can be lost, which induces super and subradiant behavior,
 an example of cooperative quantum coherent effect. 
We consider time-independent random fluctuations of the excitation
energies. This static disorder,
also called inhomogeneous broadening in literature,
 induces Anderson localization and is
able to quench Superradiance.
We identify  two different regimes: $i)$  
weak opening, in which
Superradiance is quenched at the same critical disorder at which the
states of the closed system localize; $ii)$ strong opening,  with 
a critical disorder strength proportional to both the system size
and the degree of opening, displaying robustness
of cooperativity to  disorder. 
Relevance to photosynthetic complexes is discussed.
\end{abstract}

\date{\today}
\maketitle

\section{Introduction}\label{sec:I} 

Since the discovery that quantum coherences might have a functional
role  in biological systems even at room temperature~\cite{photo,photoT,photo2,photo3,schulten}, there
has been great interest in understanding how coherences can be
maintained and used under the influence of different environments with
competing effects. In particular, much of recent research focused on one-dimensional nanostructures,
due to their relevance to molecular aggregates, such as the J-aggregates~\cite{Jaggr}, natural
photosynthetic systems~\cite{cao}, bio-engineered devices for
photon sensing~\cite{superabsorb} and light-harvesting systems~\cite{sarovarbio}.

Here we focus on  a ring-like structure of two-level systems
coupled with  nearest neighbor tunneling amplitudes which has been
recently considered  in
literature~\cite{cao,superabsorb,sarovarbio,fassioli,mukameldeph,mukamelspano}.
Usually, under low light intensity,  in many 
natural photosynthetic systems or in ultra-precise photon sensors 
the single-excitation approximation can be used. In 
this case the system is
equivalent to a tight binding model where one excitation can hop from
site to site, see Fig.~\ref{ring}.

Many photosynthetic organisms contain
ring-like chlorophyll molecular aggregates in their 
light-harvesting complexes, which are called LHI and LHII~\cite{schulten1}. These complexes have
the purpose to absorb light and to transfer the excitations to other
structures or to a central core absorber, the reaction center, where
charge separation, necessary in the
next steps of photosynthesis, occurs.
These complexes are subject to the effects of different
environments: $i)$ dissipative,   where the excitation can be
lost; $ii)$ proteic, which induces  static or dynamical disorder.
The efficiency of excitation transfer can be determined only through
a comprehensive analysis of the effects due to the interplay of all those
environments.

Here, in particular,  we consider a system subject to the influence 
of both  a common decay 
channel
where
the excitation can be lost, and a static disorder. The first
environment can be thought of  as a
 model  for the coupling of a molecular aggregate to the electromagnetic
field~\cite{mukameldeph} (loss of excitation by recombination) or for the coupling of the molecular
aggregate to a central core absorber (loss of excitation by trapping).
For many molecular aggregates, the single channel approximation is
appropriate to describe the coupling
with  the electromagnetic field,
since the wavelength of the
absorbed light is much larger than the system size (natural complexes 
such as LHI, LHII typically span few tens of nanometers, while the wavelength of the
involved photon is hundreds of nanometers).
Moreover, it can also be considered as a good approximation for the coupling to a
central core absorber, modeled for instance by a  semi-infinite one-dimensional lead~\cite{kaplan,rotterb}. 

The second environment consists of 
a protein scaffold, in which
photosynthetic complexes are embedded,  that induces 
fluctuations in the sites energies. The fluctuations which occur
on a time scale much larger than the time scale of the dynamics 
are usually described as static disorder.
By static disorder we mean position dependent, but time-independent,
fluctuations of the site energies. 
The case of time-dependent fluctuations
of site energies has been considered in a separate paper~\cite{laltro}.

It is well known that, when many sites are all coupled
to the  same channel, we can have a superradiant
behavior~\cite{Zannals}. Superradiance implies the existence of some states with a
cooperatively enhanced decay rate ({i.e.}~proportional to the number of sites). 
Superradiance comes always together with Subradiance, that is  the existence of 
states with a cooperatively suppressed decay rate ({i.e.}~smaller
than the single-site decay rate).

Though originally discovered in the context of atomic
clouds interacting with the electromagnetic field~\cite{dicke54}, in
presence of many excitations, Superradiance 
was soon recognized to be a general phenomenon in open quantum
systems~\cite{Zannals} under the conditions of coherent coupling with
a common decay channel. Most importantly,
it can also occur in presence of a
single excitation (the super of Superradiance~\cite{scully}), entailing a purely quantum effect.


The functional role that Superradiance might have in natural
photosynthetic systems 
has been discussed in many publications~\cite{schulten,superabsorb,srlloyd,sr2},
and experimentally observed in molecular aggregates~\cite{Jaggr,vangrondelle}.
Superradiance (or  Supertransfer) is also thought to play
an important role w.r.t.~the transfer of excitation to the
central core absorber~\cite{schulten},
and its effects on the efficiency of energy transport in
photosynthetic molecular aggregates have been recently analyzed~\cite{srfmo,srrc}.

The origin of Superradiance lies in the fact that the excitation can
be coherently spread over several sites, thus inducing a cooperative
effect. On the other hand, static disorder is expected to destroy Superradiance,
since it induces Localization~\cite{Anderson}, which implies that excitons
are localized on one site only, thus hindering cooperativity. 
The main question we want to address here is whether 
a critical disorder exists at which Superradiance and, thus, cooperativity
are destroyed.
The relation between Superradiance and Localization have
been already analyzed in literature
in different contexts~\cite{mukamelspano,prlT,alberto,cldipole}. 
In particular, in~\cite{mukamelspano} the case of weak coupling 
to the continuum (weak opening)  has been analyzed for one-dimensional
systems. It has been already analyzed also  by some of the Authors of the
present paper: in~\cite{alberto}  the case of open 1-d and 3-d Anderson models 
in the strong opening regime was considered.
It was pointed out there that the sensitivity to disorder
can be very different for superradiant and subradiant states:
while the latter localize at the same critical disorder of
the closed system ({i.e.}\ a system with no coupling to the continuum of states), 
the former localize at the critical disorder for
which Superradiance is quenched. 
Interestingly, though subradiant states essentially localize at the localization 
threshold associated with the closed system, they display some peculiar
 features due to opening, being  neither fully localized nor extended
(hybrid states)~\cite{alberto}. 
In this paper we aim to study both the regimes of weak and strong opening 
and their effects on Localization in one-dimensional nanostructures.

Even if  it is easy to imagine that 
opening  and disorder have competing effects on the efficiency
of energy absorption and transfer, a deeper analysis is necessary
to fully understand their action.
For instance, disorder decreases the efficiency of the 
superradiant states
in absorbing light or in transferring excitations, but, at the same time,
 it can allow for energy
absorption and transfer from the subradiant states.
Thus, for these states,
disorder is useful to enhance efficiency. 
The latter effect is strongly related to the enhancement of efficiency due 
to noise: the so called
noise-assisted transport, discussed in~\cite{lloyd,deph}.
Noise-assisted transport constitutes a general phenomenon in
quantum networks, even if its relation with Subradiance
has never been stressed up to now, to the best of our knowledge.
The plan of the paper is the following: in Sec.~\ref{sec:2} we introduce the model,
in Sec.~\ref{sec:3} we derive analytically the critical disorder strength needed to quench
Superradiance, identifying the different regimes of weak and strong opening.
In Sec.~\ref{sec:4} we analyze in detail the relation with Localization,
while Sec.~\ref{sec:5} is devoted to study 
the consequences of the previous findings on 
the system dynamics. A brief discussion about the relevance 
to photosynthetic complexes is given at the end of each Section.

\section{The Model without disorder}\label{sec:2}  
\begin{figure}[t]
\vspace{0cm}
\includegraphics[width=8cm,angle=0]{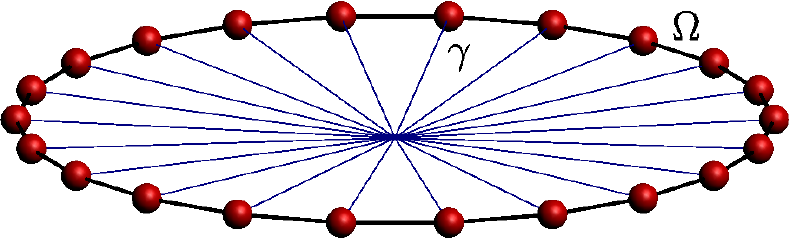}
\caption{ (Color online)
The ring model. One excitation
can hop between $N$ sites coupled with
nearest-neighbors 
 tunneling transition amplitude $\Omega$.
All sites are  connected to a common
decay channel,
where the excitation can escape,
 with an equal  coupling strength $\gamma$.
}
\label{ring}
\end{figure}

We considered a 1-d chain of sites with periodic boundary
conditions, arranged to form a ring-like structure, as shown in 
Fig.~\ref{ring}, 
where the excitation can hop from site to site.
The model is characterized by the following
tight binding Hamiltonian:
\begin{equation}
H^{tb}=  - \Omega \sum_{\langle i,j\rangle} \left(| j \rangle \langle i|
+| i \rangle \langle j|\right) \,,
\label{AM}
\end{equation}
where the summation index $\langle i,j\rangle$ runs over 
the pairs of nearest-neighbor sites
and $\Omega>0$ is the tunneling transition amplitude.
Here  $|j\rangle$ represents  a state in which the excitation is at the site $j$,
while all the other sites are unoccupied. In terms of two-level system
states ($|0\rangle, 
|1\rangle$) it can be written  as
$$ |j\rangle = |0\rangle_1 |0\rangle_2 \ldots |1\rangle_j \ldots |0\rangle_N\,. $$

The eigenvalues 
\begin{equation}
E_q= -2 \Omega \cos{\frac{2 \pi q}{N}}  \text{ with }  q=1,\ldots,N
\label{eigq}
\end{equation}
and the eigenstates $|\psi_q\rangle$ of the
system can be computed exactly.
Concerning the components of the  eigenstate $|\psi_q\rangle$   on the site basis $|s\rangle$,
one has
\[
\braket{s}{\psi_q}=\frac{1}{\sqrt{N}} \cos { \frac{2 \pi sq}{N}}
\]
for $q=1,\ldots,N/2,N$, and
\[
\braket{s}{\psi_q}=\frac{1}{\sqrt{N}} \sin { \frac{2 \pi s(N-q)}{N}}
\]
for $q=N/2+1,\ldots,N-1$.
The ground state, corresponding to  $q=N$ and energy $ E_{N}=-2 \Omega $, is fully symmetric and extended in the site basis:
\begin{equation}
|\psi_N \rangle=\frac{1}{\sqrt{N}} \sum_{k=1}^N |k\rangle.
\label{SR}
\end{equation}

The 1-d Anderson model can be ``opened'' by allowing
the excitation  to escape the system from any site into the same continuum
channel. This situation of ``coherent dissipation'' can be met
in many systems and it has been recently considered in~\cite{alberto},
where it has been shown to give rise to the following 
effective non-Hermitian Hamiltonian (see also~\cite{rotterb}): 
\begin{equation}
(H_{\mathrm{eff}})_{ij}=(H^{tb})_{ij} -\frac{i}{2}  \sum_c A_i^c (A_j^c)^* \equiv
 (H^{tb})_{ij} -i\frac\gamma2 Q_{ij},
\label{amef}
\end{equation}
where 
$A_i^c$ are the transition amplitudes
from the discrete state $i$ to the continuum channel $c$.
In our case, we have a single decay channel, $c=1$, and   equal
 couplings, $A_i^1= \sqrt{\gamma}$,  so that $Q_{ij}=1$ $\forall i,j$.

The quantum evolution is  given by the operator $${\cal U } = e^{-iH_{\mathrm{eff}}t/\hbar}\,,$$
which is non-unitary, and gives rise to a loss of probability in the decay channel.
The complex eigenvalues of $H_{\rm eff}$ can be written as
$E_r -i \Gamma_r/2$, 
where  $\Gamma_r$ represent the decay widths of the eigenstates. 
Usually, in molecular aggregates, energy is measured 
in units of $\textrm{cm}^{-1}$, corresponding to energy divided by $hc$. In these units, time is measured in $\textrm{cm}$ which corresponds to  the mapping $t \to 2\pi c t$  ($ c \simeq 0.03 \ \textrm{cm/ps}$ is the speed of light).
In the following all units of energy will be given in $\textrm{cm}^{-1}$ and in order 
to have time in $\textrm{ps}$ we need to divide it by  $ 2\pi c $.

Due to its specific structure, the operator $Q$ has only one eigenstate
with a non-zero eigenvalue:  this is the fully extended state
with eigenvalue equal to $N$. 
This eigenstate also corresponds to the
ground state of $H^{tb}$, given in Eq.~(\ref{SR}).
All the other eigenstates of $Q$ are degenerate with null eigenvalue and, since $[Q, H^{tb}] = 0$, 
they can be chosen to match the eigenstates $|\psi_q\rangle$, $q<N$, of $H^{tb}$.
 This implies that only the state
$|\psi_N\rangle$, Eq.~(\ref{SR}), has a non-vanishing decay width
equal to the total decay width of the system: 
$\Gamma_N=N\gamma$. This is the superradiant state.
Note that the  dependence on $N$  of that decay width is the
hallmark of the cooperative nature of Superradiance.
All the other states with zero decay width 
are called subradiant. The full expression for the complex eigenstates
of the non-Hermitian Hamiltonian, Eq.(\ref{amef}), is given in
Appendix $A$, see Eq.(\ref{ceig}).
Importantly, the superradiant effect might explain the
strong dependence on the initial state of the efficiency of energy
transfer to a central core absorber discussed in
Ref.~\cite{fassioli}.

Several features of the  model above in absence of
disorder  are quite atypical. Indeed, Superradiance, as discussed in many
papers~\cite{puebla,Zannals}, 
 is usually reached only above a critical coupling strength with the continuum (in the
 overlapping resonance regime) when
\begin{equation}
\label{orc}
\langle \Gamma \rangle/\Delta \ge 1,
\end{equation}
 where $\langle \Gamma \rangle$ is the average decay width and $\Delta$ is the
 mean level spacing of the closed system.
On the other hand,  
{\it    we are in a superradiant regime for any $\gamma >0$}, 
even if the overlapping resonance condition is not satisfied.
Moreover,  the widths of the subradiant states are 
usually small, but not zero as in this case.
This is also due to the particular symmetric configuration chosen, from which
it follows that $H^{tb}$ and $Q$ commute.
Note that such geometrically-induced subradiant subspaces with
zero decay width are equivalent to the trapping-free subspaces
discussed in literature~\cite{deph}.

Finally, let us notice that 
the presence of a superradiant regime for any coupling strength to the continuum 
might indicate a relation between
structure and function in natural complexes, and it might also suggest the use of
ring-like structures to exploit the superradiant behavior.

\section{Superradiance and analysis of decay widths in presence of diagonal disorder}\label{sec:3}

The peculiar features discussed above disappear when we introduce the  diagonal
disorder, described by adding to the Hamiltonian,   Eq.~(\ref{amef}), the term
\begin{equation}
D =  \sum_{j=1}^{N} \epsilon_j | j\rangle \langle j|,
\label{dam}
\end{equation}
where the random diagonal energies
$\epsilon_j$ are taken uniformly distributed in $[-W/2 ,+W/2]$, $W$ being  the disorder strength.
With the addition of this term our model becomes equivalent to a 1-d
open Anderson model as considered in Ref.~\cite{alberto}. 

The presence of static disorder on the site energies
breaks the symmetry of the system under rotations, inducing the width of the
superradiant state to decrease and the widths of the subradiant states
to increase (the total decay width $N\gamma$ 
is a constant that does not depend on the degree
of disorder $W$), so that all of the eigenstates can decay into the continuum channel.

\begin{figure}[t]
\vspace{0cm}
\includegraphics[width=8cm,angle=0]{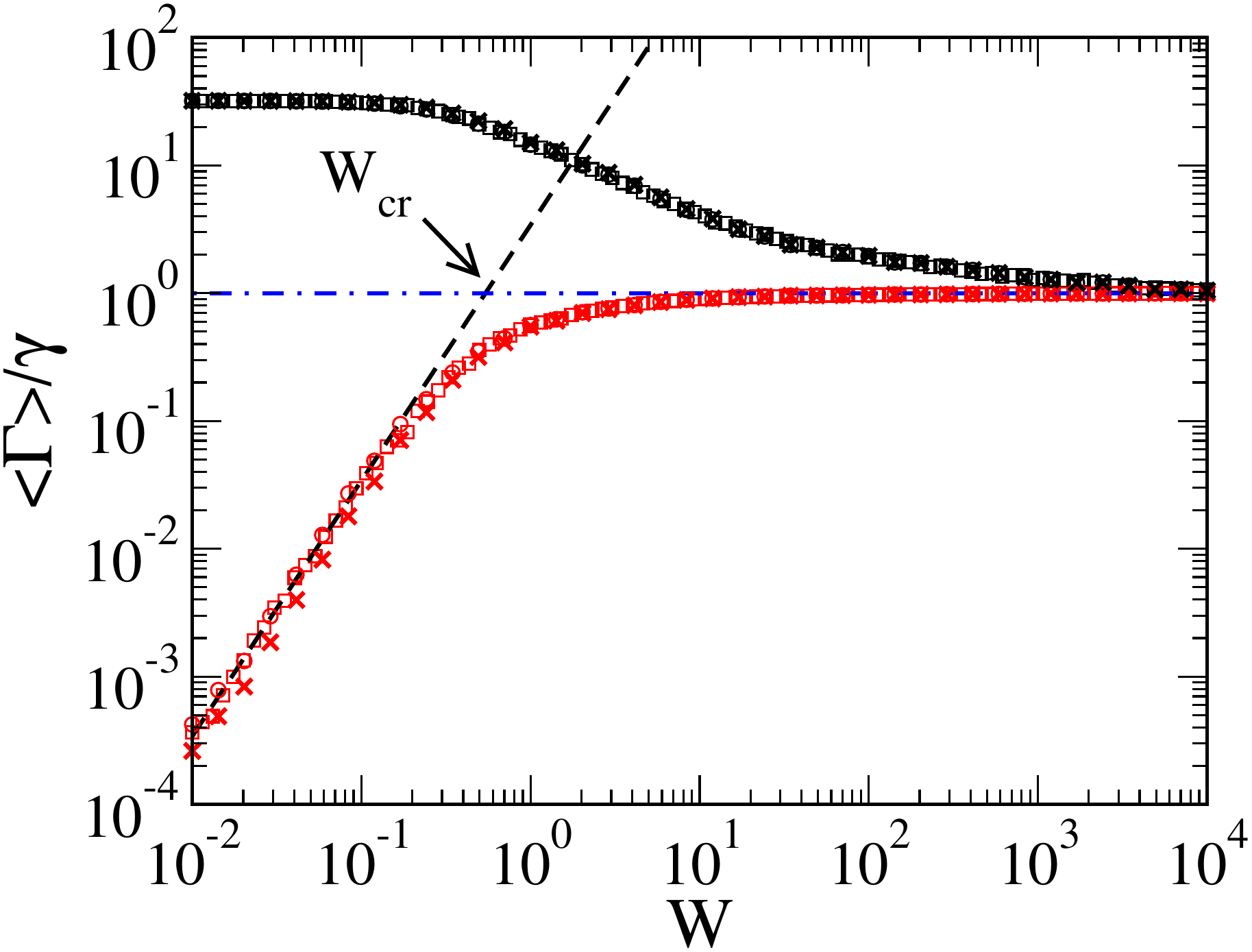}
\caption{ (Color online)
Average decay widths for a ring with $N=32, \Omega=1$
 {\it vs} the disorder strength, $W$, for $\gamma=10^{-4}$ (circles), $\gamma=10^{-3}$ (squares), 
 and $\gamma=2 \times 10^{-3}$
 (crosses). 
Black symbols (upper curve) refer to the average over disorder of the
  largest width
(the superradiant state).
Red symbols (lower
curve) refer to the average over disorder of the mean subradiant width (mean
taken over the $N-1$  smallest widths)
$\langle \Gamma_{sub} \rangle$.
As a dashed line we plot  
the perturbative result, Eq.~(\ref{Gsub}), for $\langle \Gamma_{sub}
\rangle$.
The horizontal dot-dashed line indicates the value $\langle \Gamma
\rangle=\gamma$.
The critical disorder strength, Eq.~(\ref{Wlimit}),  is indicated as  the
intersection between  the  line given by perturbation theory 
and the horizontal line, see
text for details.
}
\label{GW}
\end{figure}

The effect of static disorder on the decay widths has
been  analyzed in  
Fig.~\ref{GW}, where the width of the
superradiant state and  the average width of all  subradiant states
are shown as a function of the disorder strength $W$, for different parameters value. 
As one can see, for small disorder, the effect on subradiant states is much
more evident than that on the superradiant state. 
For large disorder, all  widths approach the value $\gamma$, corresponding
to the decay width of an isolated site. In this regime there is no
collective behavior anymore and Superradiance is completely quenched.

For small disorder, 
it is possible to  use  perturbation theory (see Appendix~\ref{app:A})
to obtain the mean decay width of the $N-1$ smallest widths:
\begin{equation}
\begin{aligned}
\langle \Gamma_{sub} \rangle &\mbox{}=\frac{\gamma W^2}{48\Omega^2(N-1)}\\
&\mbox{}\times
\sum_{q=1}^{N-1}
\left[\left(\cos\frac{2\pi q}{N}-1\right)^2+\frac{N^2\gamma^2}{16\Omega^2}\right]^{-1}\,.
\end{aligned}
\label{Gsub}
\end{equation}
The sum in Eq.~(\ref{Gsub}) can be well
approximated in different parameter regimes to give
(see Appendix~\ref{app:B})
\begin{equation}
\langle{\Gamma}_{sub}\rangle  = \left\{
\begin{array}{lll}
\displaystyle \frac{ \gamma W^2 N^3}{ 96 \pi^4 \Omega^2}  \quad &{\rm for} \; 
&\displaystyle   \frac{N \gamma }{2 } \ll  \delta E_{min}\,,  \\
   &   & \\
\displaystyle \frac{ \gamma^{1/2} W^2 }{ 12  \Omega^{1/2}N^{3/2}   }  \quad &{\rm for} \; 
&\displaystyle  \delta E_{min} \ll \frac{N \gamma }{2 }  \ll 2 \Omega\,,  \\  
   &   & \\
\displaystyle \frac{ W^2 }{ 3 N^2 \gamma    }  \quad &{\rm for} \;
&\displaystyle   \frac{N \gamma }{2 }  \gg  2 \Omega\,.
\end{array}
\right.
\label{Gsub2}
\end{equation}

Let us name \emph{weak opening} the regime characterized by $N \gamma /2 \ll  \delta E_{min}$
and \emph{strong opening} the one in which $N \gamma /2  \gg  2 \Omega $.
The different regimes shown above can be understood if we consider that 
 $$\delta E_{min}=E_{N-1}-E_{N} \simeq 4 \Omega \pi^2/N^2$$ 
is the
minimal nearest neighbor energy distance, see 
Eq.~(\ref{eigq}). 
In Ref.~\cite{mukamelspano} a perturbative result was obtained in the regime of  weak opening,
$N \gamma /2\ll  \delta E_{min}$, which agrees with our findings.
As one can see from Eq.~(\ref{Gsub2}),
the average subradiant width  in any regime
increases as $W^2$, but the dependence on the system size, $N$,
and on the degree of opening, $\gamma$,
changes: in the weak opening regime, the widths
increase with $N$ and $\gamma$, whereas, for very strong opening, they decrease
with $N$ and $\gamma$.

In  Fig.~\ref{GW} the perturbative expression is shown as a dashed
line and agrees very well with numerical data. From Eq.~(\ref{Gsub})
one can define a critical disorder strength, $W_{cr}$, at which
Superradiance is quenched, given by the condition 
\begin{equation}
\langle \Gamma_{sub} (W_{cr})\rangle=\gamma,
\label{defw}
\end{equation}
 from which one gets
\begin{equation}
W_{cr}=\sqrt{
\frac
{48\Omega^2(N-1)}
{
\sum_{q=1}^{N-1}\left[\left(\cos\frac{2\pi q}{N}-1\right)^2
+\frac{N^2\gamma^2}{16\Omega^2}
\right]^{-1}
}
}\,.
\label{Wcr}
\end{equation}

For $W \gg W_{cr}$, all of the widths become essentially the same 
and equal to $\gamma$, while below $W_{cr}$
they strongly depend on the chosen state. 
Usually, the transition between these two regimes, which corresponds also to a transition from a non-cooperative to a cooperative regime, is referred to as 
Superradiance Transition (ST) in literature~\cite{Zannals,puebla}.

Even if  the validity of Eq.~(\ref{Wcr}) has been shown in Fig.~\ref{GW} only
in the weak opening  regime, 
we checked that it gives an excellent estimate of the disorder
 at which Superradiance is quenched also for strong opening.

From Eq.~(\ref{Wcr}) it is possible to 
get an approximate expression (see Appendix~\ref{app:B}) for the critical disorder strength, 
$W_{cr}$, in the different regimes:
\begin{equation}
W_{cr} = \left\{
\begin{array}{lll}
\displaystyle \sqrt{96} \pi^2 \Omega N^{-3/2} & \quad {\rm for} \; 
&\displaystyle    \frac{N \gamma }{2 } \ll  \delta E_{min}\,,  \\
  &    & \\
\displaystyle \sqrt{12} (\gamma \Omega N^3)^{1/4} & \quad {\rm for} \; 
&\displaystyle   \delta E_{min} \ll \frac{N \gamma }{2 }  \ll 2 \Omega\,,  \\  
&      & \\
\displaystyle \sqrt{3}  N \gamma & \quad {\rm for} \;
&\displaystyle    \frac{N \gamma }{2 }  \gg  2 \Omega\,.  \\      
     \end{array}
\right.
\label{Wlimit}
\end{equation}
The results contained in  Eq.~(\ref{Wlimit}) are very interesting, since they show that in some region of parameters (typically small system size and weak opening)
the critical disorder at which  Superradiance is quenched
is independent of $\gamma$ (a quantity often difficult
to be experimentally determined),
while it decreases with the system size as $N^{-3/2}$. This
independence from $\gamma$ is also shown in Fig.~\ref{GW} where
we plotted data obtained with
different values of $\gamma$, for which $N\gamma/2 \le \delta
E_{min}$.  In particular, for the largest value of $\gamma$ considered in
Fig.~\ref{GW}, $\gamma=2\times10^{-3}$, we have $N\gamma/2 \simeq \delta
E_{min}$. 
The existence of a regime (weak opening)  in which the critical disorder strength
is independent of $\gamma$ could be surprising. Indeed, applying
the overlapping resonance criterion, Eq.~(\ref{orc}),
one would obtain $W_{cr} \propto \gamma$,
since $\langle \Gamma \rangle \propto \gamma$
and  $\Delta \propto W$.
An explanation of this effect, due to Localization, will be given in the next Section.


A second remarkable result is the linear dependence of $W_{cr}$ on $N$ and $\gamma$
 in the strong opening regime.
Since, on increasing $N$, one always enters the strong opening regime, it is possible to preserve the cooperative nature
of superradiant states up to arbitrarily large disorder.
We may thus say that the opening
induces a cooperative robustness to disorder,  
 as was also 
recently found by some of the Authors of this paper~\cite{alberto}.

It is interesting to observe that 
 also in the case of dynamical disorder 
it was found~\cite{laltro} that the critical dephasing necessary to destroy the 
cooperative superradiant effects is proportional to both $N$ 
and  $\gamma$.
Note that this regime was not analyzed in Ref.~\cite{mukamelspano}, where it was stated 
that the critical disorder
needed to quench Superradiance does not depend on the superradiant 
decay rate. 

From Eq.~(\ref{Wlimit}) we can infer that
the dependence of $W_{cr}$ on 
 the system size $N$ 
is non-monotone. Setting $N\gamma/2 = \delta E_{min}$, we can
roughly estimate  the $N$ value
at which $W_{cr}$ has a minimum:
\begin{equation}
N_{cr} \simeq \left( \frac{8 \pi^2 \Omega} {\gamma} \right)^{1/3}.
\label{Ncr}
\end{equation} 

\begin{figure}[t]
\vspace{0cm}
\includegraphics[width=8cm,angle=0]{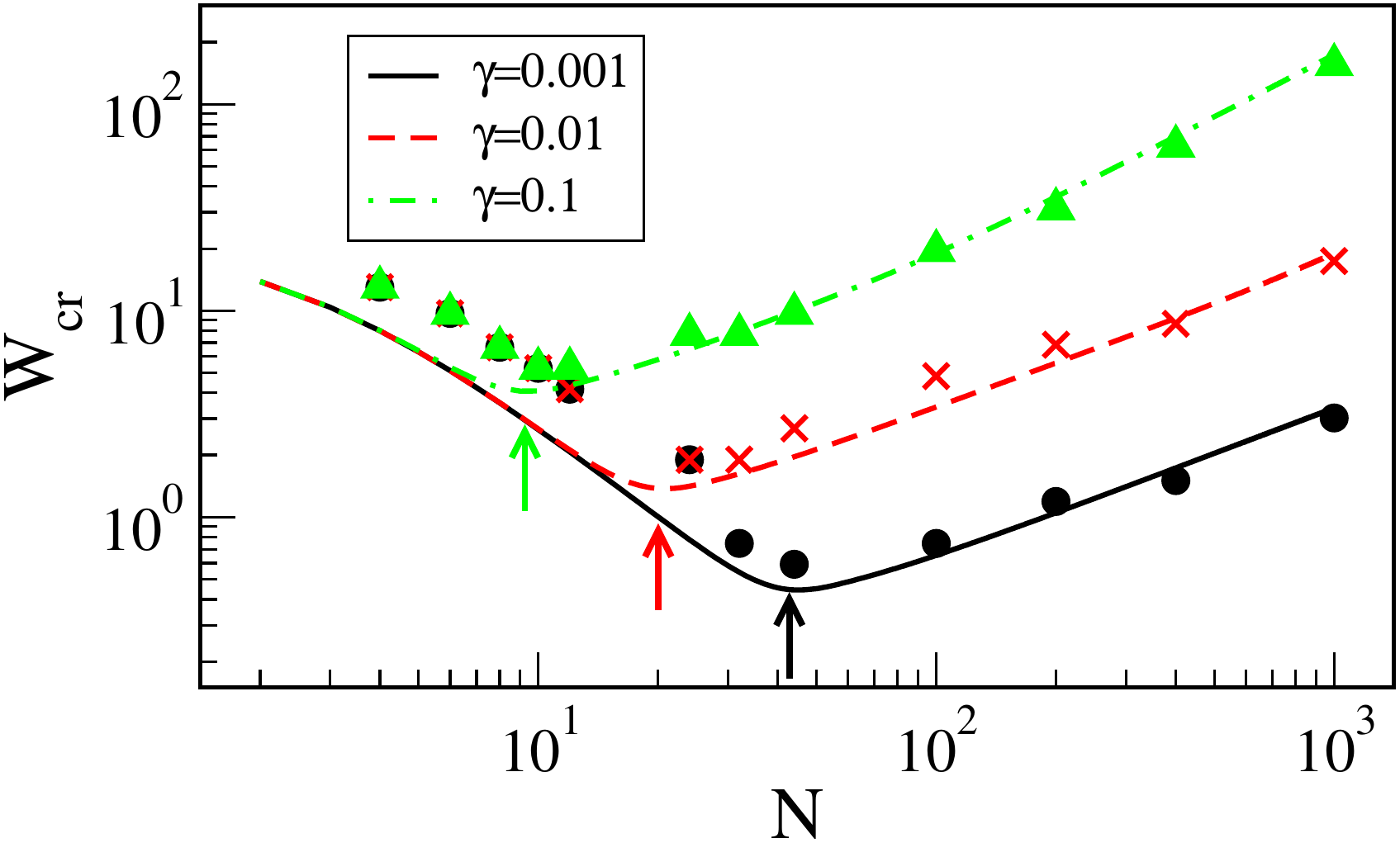}
\caption{ (Color online)
Disorder strength at which the variance of the decay widths have a
maximum  {\it vs} the number $N$ of sites in the ring for
different values  of the coupling strength $\gamma$, as indicated in the legend.
The numerical data (symbols) are compared with the analytical expression 
(curves) for the
critical disorder strength, Eq.~(\ref{Wcr}). Arrows indicate the
size $N_{cr}$ at which critical disorder is minimal, see Eq.~(\ref{Ncr}).
}
\label{VarG}
\end{figure}

To confirm the validity  of the critical disorder strength, $W_{cr}$,
 computed above, as the value at which the ST occurs, we computed the variance of the decay widths.
Indeed, it is well known~\cite{puebla} that, at the ST, the variance of
the widths has a maximum. %
The results of such a comparison are presented in Fig.~\ref{VarG}, showing a good agreement between the two estimates of the ST.

Finally, it is interesting to estimate the value of $W_{cr}$ for the
photosynthetic complexes LHI and LHII. In this case $\Omega \approx
600\,\mathrm{cm}^{-1}$ and $N=32, \ 16$, respectively~\cite{schulten,schulten1}.
The  coupling with the electromagnetic field can be estimated from the
radiative decay time $\tau$  of a single molecule, which is of the
order of few nanoseconds~\cite{schulten1}. Hence we get $\gamma= 1/(2
\pi c \tau) \approx 10^{-3}\,\textrm{cm}^{-1}$. On the other side, for
the LHI complex, the common decay channel can also represent the
reaction center. This coupling  can be estimated to be $\gamma \approx 10^{-2}\,\textrm{cm}^{-1}$ from the mean transfer time to the reaction center of the LHI complex~\cite{schulten1}, as discussed at the end of Sec.~\ref{sec:5}.
Both these couplings are very weak
if compared to the energy scale of $\Omega$, so that we can assume that we
are in a weak opening regime, $N \gamma/ 2  \ll  \delta E_{min}$, where $W_{cr}$ does not depend on $\gamma$ and it will then be the same for both environments. 
We can thus use $W_{cr}=\sqrt{96} \pi^2 \Omega N^{-3/2}$,
see Eq.~(\ref{Wlimit}), getting  $W_{cr} \approx 320\,\mathrm{cm}^{-1}$ for LHI and $W_{cr} \approx 900\,\mathrm{cm}^{-1}$ for LHII. 
These values of disorder are in agreement with the
experimental observation that static disorder in LHII complexes is
$2$--$3$ times larger than the value of disorder in LHI complexes~\cite{vangrondelle}.
Those values are also quantitatively compatible with the
estimated ranges of static disorder strength in natural photosynthetic
complexes ($100$--$600\,\mathrm{cm}^{-1}$ for LHI complexes~\cite{vangrondelle,fassioli}, 
$600$--$1400\,\mathrm{cm}^{-1}$ for LHII
complexes~\cite{vangrondelle,schulten,schulten1}. To make a comparison
with the data contained in these references, one should take into account that
they considered 
Gaussian static disorder with a standard
deviation $\sigma$, so that 
$W=\sqrt{12} \sigma$).
These estimates  might suggest that natural
photosynthetic complexes operate close to the ST.


\section{Superradiance and localization}\label{sec:4}

In the previous Section we analyzed how diagonal disorder
modifies the decay widths of the states. On the other hand, it is well
known that disorder in isolated tight binding models induces Anderson
Localization~\cite{Anderson}. In  1-d
systems any disorder strength induces localized eigenstates,
$|\langle j |\psi \rangle | \simeq  \exp(-|j-j_0|/\xi)$,  
where $j$ labels the position of the sites on the lattice  and  $\xi$ is the 
localization length, measured in units of intersite distance. The
localization length  is, in general, a function 
of the disorder strength $W$ and of the energy $E$.
In particular, it is well known that, for weak disorder 
and away from  the
edges of the energy band,  $\xi \propto W^{-2}$.

Therefore,  it is possible to define a critical
disorder $W_d$ for the localization effect to be important by
the simple equation
\begin{equation}
\xi (W_d)  = N\,.
\label{loc88}
\end{equation} 
Indeed, while any  
increase of the disorder strength will produce eigenstates 
with a localization length smaller than the sample size, decreasing $W$ gives
rise to eigenstates with a localization length larger than the system size, {i.e.}\ effectively delocalized.

\begin{figure}[t]
\vspace{0cm}
\includegraphics[width=8cm,angle=0]{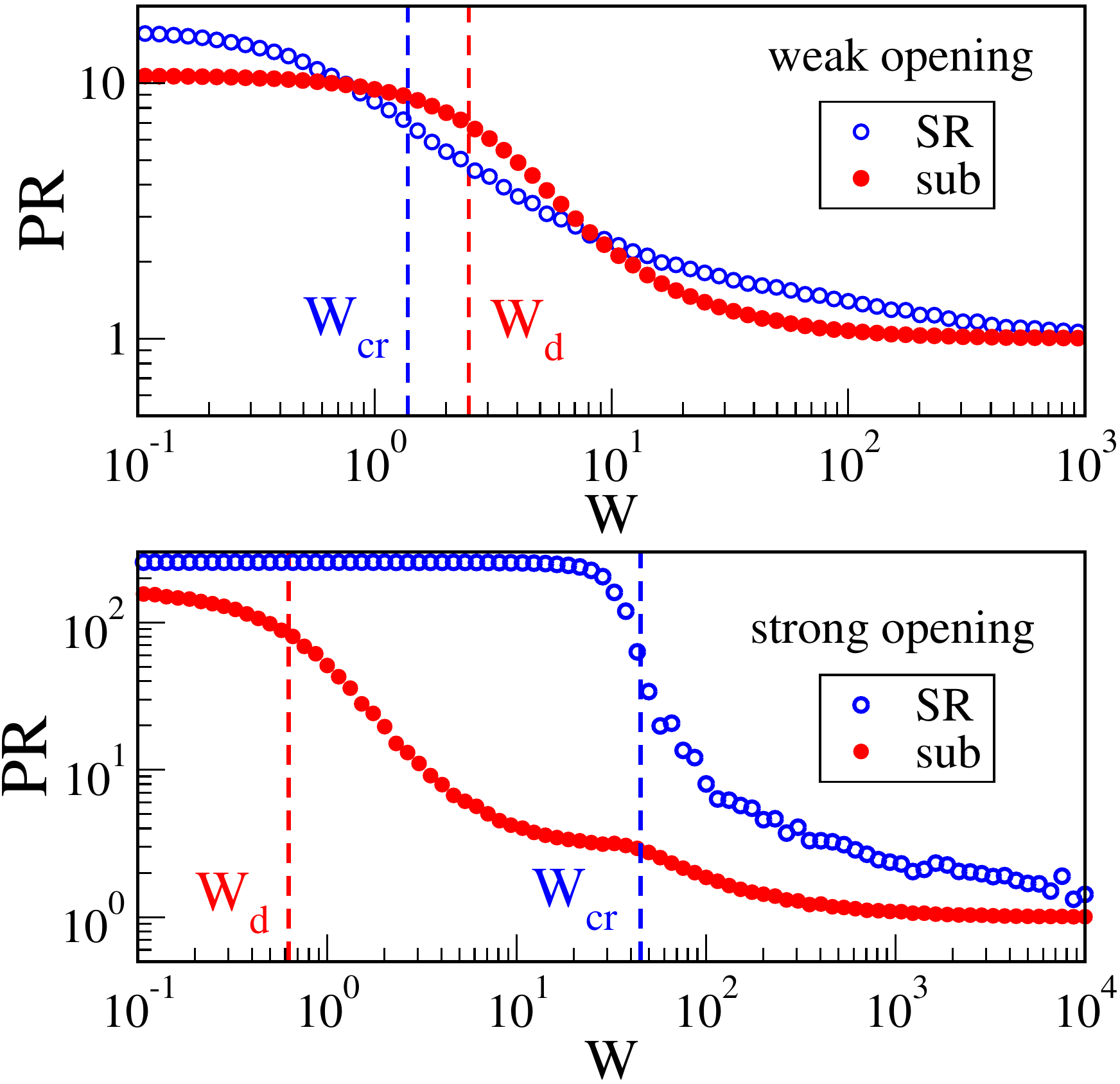}
\caption{ (Color online)
Average participation ratio $PR$  {\it vs} the disorder strength
$W$. Upper panel represents the
weak opening regime, namely $N=16, \gamma=10^{-3}, \Omega=1$,
while the lower panel depicts the strong opening regime,
$N=256, \gamma=10^{-1}, \Omega=1$.
In both panels the blue open circles represent the behavior
of the superradiant state as a function of the disorder strength,
while the red full circles stand for the average $PR$ of the subradiant states.
Vertical dashed lines represent in both panels the localization condition given by Eq.~(\ref{loc88}) (red), and $W_{cr}$  (blue, Eq.~(\ref{Wlimit})).
}
\label{PRring}
\end{figure}

The interplay of disorder and opening can be studied by means of the participation ratio
\begin{equation}
 PR= \left\langle              
\frac{ \sum_i |\langle i| \psi \rangle|^2  }
{ \sum_i |\langle i| \psi \rangle|^4  }
\right\rangle
 \label{pr}
\end{equation}
of the eigenstates $|\psi \rangle$  of $H_{\mathrm{eff}}$, 
given in Eq.~(\ref{amef}),
where  $\langle {\dots} \rangle$  stands for the ensemble average over 
different realizations of the static disorder.
The $PR$ is widely used to characterize localization properties~\cite{pr}
and it clearly satisfies the bounds $1\le PR\le N$.
For extended states, it increases proportionally to the system size $N$,
while, for localized states, it is independent of $N$.

Our aim is to compare the disorder strength at which Superradiance is
quenched, $W_{cr}$, see Eq.~(\ref{Wcr}),
with the disorder strength  at which the states localize, $W_d$, see Eq.~(\ref{loc88}).
To do that, we analyze separately the $PR$ of the superradiant state
(the state with the maximum decay width) and the average $PR$ of the
other $N-1$ states as a function of  $W$.

The typical behavior of the $PR$
as a function of the disorder strength $W$ has been analyzed in Fig.~\ref{PRring}
in two different situations
: for weak opening ($N \gamma/2 \ll 
\delta E_{min} $, upper panel), and for strong opening ($N \gamma /2 \gg 2 \Omega$, lower panel).
In both cases, the $PR$
of the superradiant state decreases roughly at the ST, as given by $W_{cr}$,
while the $PR$ of  subradiant states decreases roughly at $W_d$.

\begin{figure}[t]
\vspace{0cm}
\includegraphics[width=8.5cm,angle=0]{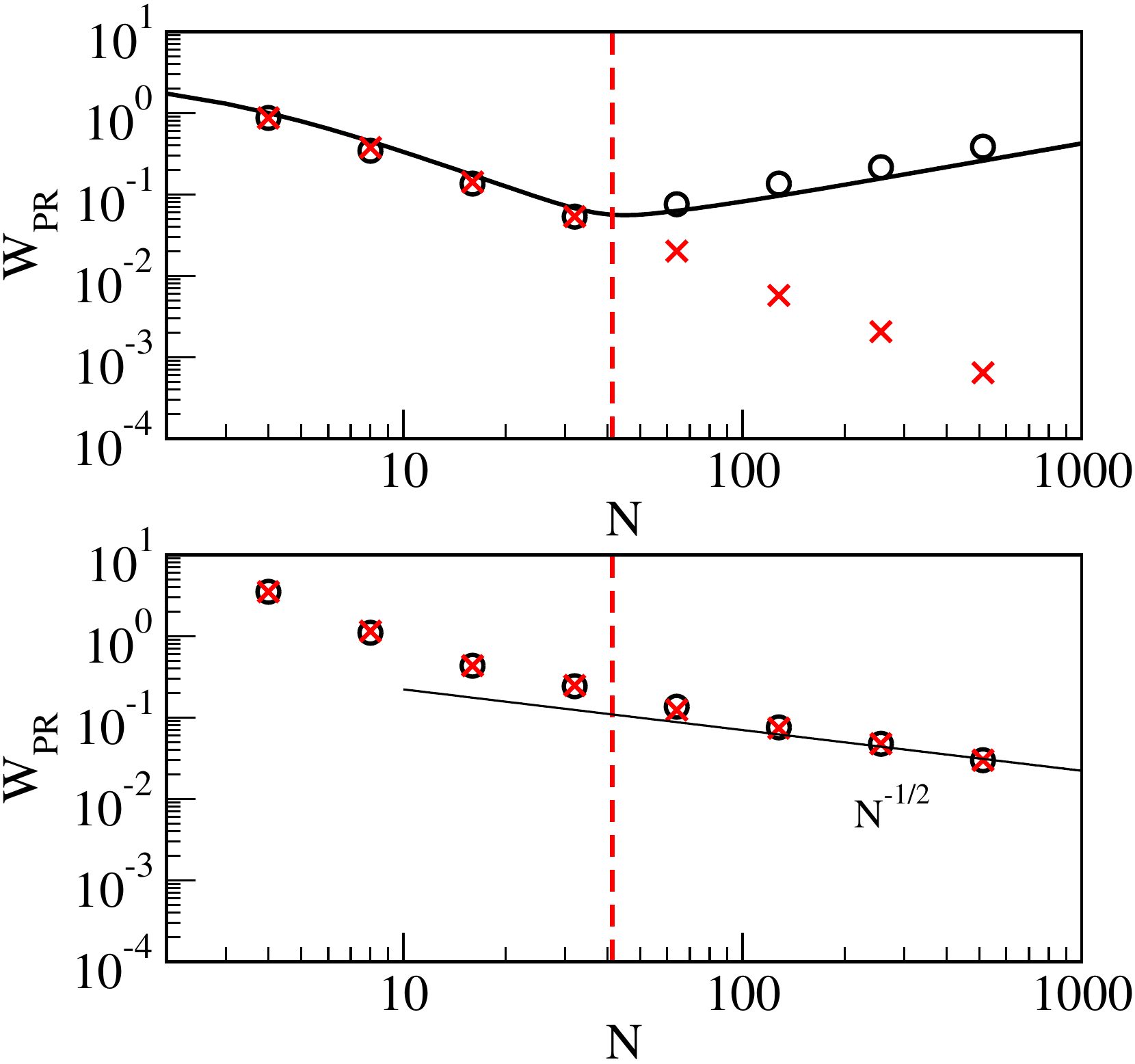}
\caption{ (Color online)
Disorder strength at which the $PR$
decreases its value by $3\%$ with respect to the value at $W=0$,
as a function of the system size, $N$,
for the superradiant state (upper panel) and
for the average $PR$ over all the other states (lower panel).
Circles stand for the open system ($\gamma=0.001$),
crosses for the  closed system ($\gamma=0$).
Here is $\Omega=1$.
In the upper panel the full curve stands
for the analytical $W_{cr}$, given by Eq.~(\ref{Wlimit}),
 rescaled by a factor $8$ to fit numerical data.
In the lower panel a curve proportional to $N^{-1/2}$ has been drawn 
to guide the eye (for the explanation see text). 
Vertical dashed lines mark $N_{cr}$, see Eq.~(\ref{Ncr}), and separate the weak opening regime (left) from the strong opening regime (right).
}
\label{WcrPR1}
\end{figure}

To be more quantitative, we numerically computed, for 
the superradiant state and for the subradiant states, the disorder strength
$W_{PR}$ at which their $PR$ decreases by 3\% w.r.t.~the value at zero disorder.
To highlight the peculiar effects due to opening, these
results should be compared with those for the closed system ($\gamma=0$). 
For the closed system we cannot define superradiant and subradiant states, but, 
since the localization length depends on the energy level, 
we can compare states of the open system with states of the closed system
having the same real energy.
In particular, the superradiant state is compared with the ground state of the closed system.

Results are shown in Fig.~\ref{WcrPR1} for the superradiant state (upper panel) 
and for the subradiant ones (lower panel) as a function of the system size $N$.
In this Figure we fix $\gamma$  and, by varying $N$, we switch from the weak
opening regime (for small $N$ values) to the strong opening regime (for large $N$).
The $N_{cr}$ value which separates the two different regimes
 can be estimated from  Eq.~(\ref{Ncr}), and has been indicated as a dashed 
vertical line in both panels.

The opening  does not modify the behavior of
the subradiant states  if compared with the behavior of 
the closed system, compare  circles with crosses in the lower panel
of Fig.~\ref{WcrPR1}.
In particular, for non-edge states of the closed system~\cite{felix}
$\xi \simeq 100/W^2,$
and, from Eq.~(\ref{loc88}),  one gets
that the disorder strength at which states localize scales as
$ N^{-1/2}.$
The same dependence on $N$  is found in presence of opening and it has been
indicated for the sake of comparison in  Fig.~\ref{WcrPR1}, lower panel.

Let us now analyze the behavior of superradiant states, Fig.~\ref{WcrPR1}
upper panel. 
In the weak opening regime, $N < N_{cr}$, the open and the closed model
display the same behavior, while in the strong opening regime, $N > N_{cr}$,
the behavior is very different:
in this regime 
 $W_{PR}$ decreases with $N$ for the closed model, while it 
increases with $N$ for the open one.

Even if the behavior of the superradiant state of the open system 
in the weak and strong opening regimes is very different, 
it is always  captured by 
the disorder strength at which Superradiance is quenched,
$W_{cr}$, see Eq.~(\ref{Wlimit}).
Indeed, 
the disorder strength  at which
 the superradiant state starts lo localize (phenomenologically
described by $W_{PR}$)  scales with the parameters
as $W_{cr}$
(compare full line with symbols in upper panel of Fig.~\ref{WcrPR1}).
This fact allows us to understand the scaling of $W_{PR}$ with $N$ in
both regimes: $W_{PR} \propto N^{-3/2}$ in the weak opening regime,
while $W_{PR} \propto N$ in the strong opening regime.

Note that the dependence $W_{PR} \propto N^{-3/2}$
 is the same as that of the
disorder strength necessary to localize the edge states of the closed
system~\cite{PRBloc}, for which we have $\xi(W) \propto W^{-2/3}$
and, from Eq.~(\ref{loc88}),
we obtain a disorder strength scaling as $N^{-3/2}$.

The different sensitivity of super and subradiant states
to disorder is far from being trivial. Due to the fact that the $Q$
matrix in Eq.(\ref{amef}) is a full matrix, the opening induces a long-range hopping which contrasts
localization, and one might expect such  long range to affect all
states equally. On the other hand, the correlated nature of the long
range hopping implies that only superradiant states are affected,
leaving the subradiant states effectively decoupled from the
environment and thus behaving more similarly to the states of the closed system. For
more details see Ref.~\cite{alberto}.

Summarizing  we can conclude that:
\begin{itemize}
\item  the  disorder  strength 
necessary to localize the subradiant states
is the same of the corresponding value
for the closed system;

\item the  disorder  strength 
necessary to localize the superradiant states is proportional to the
disorder strength necessary to quench Superradiance, $W_{cr}$;

\item in the   weak opening regime,
the quenching of Superradiance is determined only by the Localization properties of the
closed model, resulting in a  $W_{cr}$   independent of $\gamma$;

\item in the strong opening regime, Superradiance is quenched
at a critical disorder proportional to $N\gamma$, as 
in the case of time-dependent disorder~\cite{laltro}, thus showing 
its cooperative robusteness to disorder;

\item for the realistic parameters of natural
  photosynthetic complexes, such as LHI and LHII (see end of
  Sec.\ \ref{sec:3}), we are in the weak opening regime, so that it is possible
  to determine $W_{cr}$ only by analyzing the localization properties
  of the closed system. This fact can be very useful since the exact value
  of $\gamma$ is not easy to be determined experimentally.
\end{itemize}


\section{Dynamics of the Survival Probability}\label{sec:5} 

\begin{figure}[t]
\vspace{0.cm}
\includegraphics[width=8cm,angle=0]{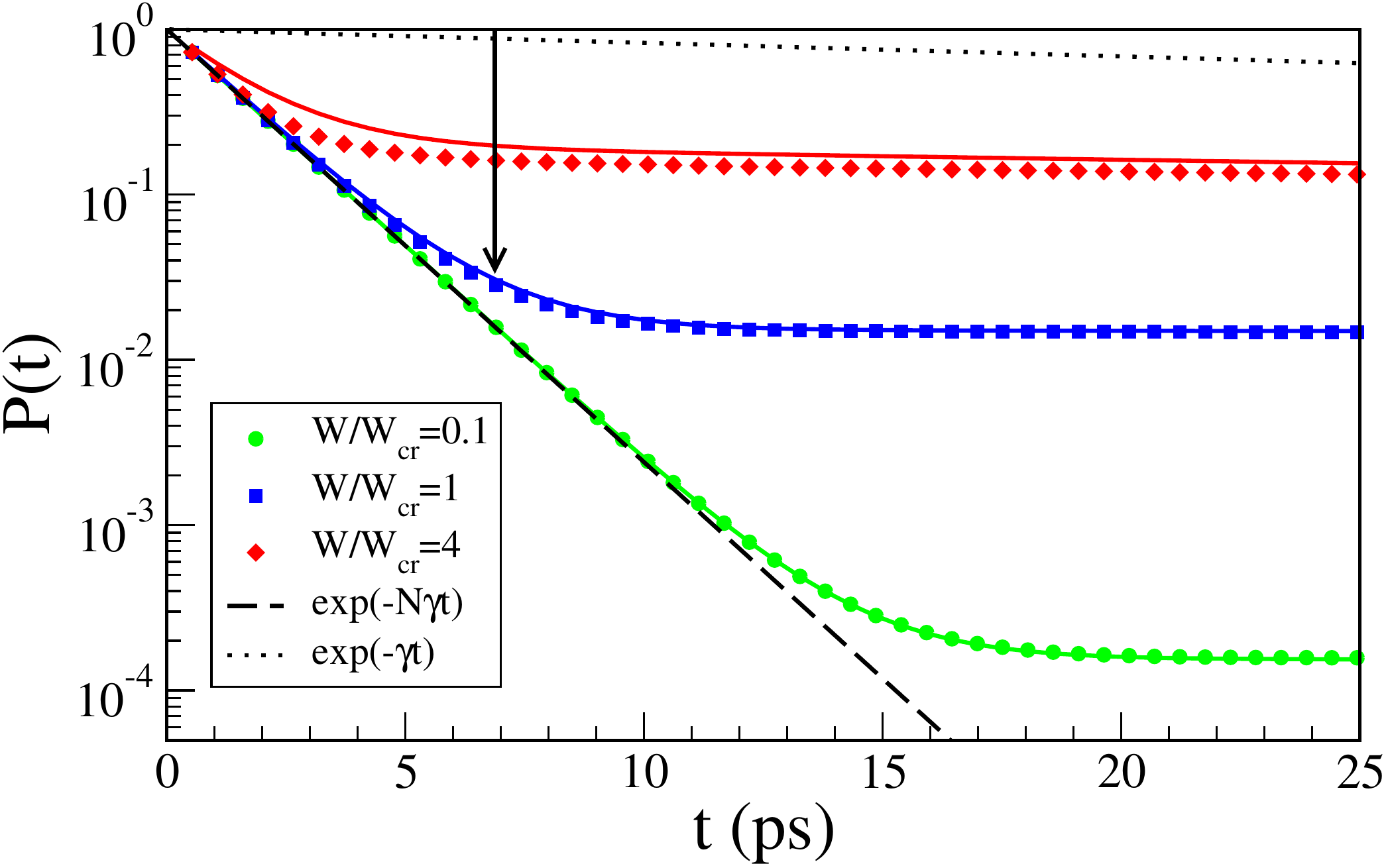}
\caption{(Color online)
Time evolution of the survival probability starting from the fully extended state. We set
$N=32$, $\gamma=0.1$, $\Omega=1$, for different disorder strengths
as indicated in the legend.
Numerical results are shown with symbols, while full curves
show the analytical expression, Eq.~(\ref{Pth}). The arrow shows the
value of $t^{\star}$ obtained analytically by Eq.~(\ref{ts}) for
$W=W_{cr}$. For comparison, both
the exponential decay of the superradiant state for zero disorder (dashed) and
the decay at large disorder (dotted) are shown.
}
\label{Ptth}
\end{figure}

In this Section we aim at studying how the time evolution of
the survival probability $P(t)$ (that is the probability of finding the
excitation in the system, initially prepared in some 
 state $|\psi_0 \rangle$) is modified by the presence of static disorder.
That quantity is given by
\begin{equation}
\label{pstay}
P(t) = \sum_{k=1}^N | \langle k | e^{-iH_{\mathrm{eff}}t/\hbar} |\psi_0 \rangle |^2.
\end{equation}

Let us choose $|\psi_0 \rangle=|\psi_N \rangle$, the fully extended  
state of Eq.~(\ref{SR}), for our first analysis.
For $W=0$ we clearly have
$P(t)=e^{-N\gamma t}$, since the fully extended state is the only one
with a decay width, see Eq.(\ref{ceig}). 
For $W \ne 0$ the fully extended state 
does not coincide with the superradiant state anymore, and it 
should be written as a superposition of superradiant and subradiant
states. 
Using first order perturbation theory (in the disorder strength $W$)
we can derive an approximate expression for $P(t)$ valid for small time, see Appendix~\ref{app:A}, Eq.~(\ref{c1}): 
\begin{equation}
P(t) \approx c_1 e^{-N\gamma t} + (1-c_1) e^{-\Gamma_{sub}^{max} t}\,.
\label{Pth}
\end{equation}
Eq.~(\ref{Pth}) takes into account only the
superradiant decay and the fastest subradiant decay, $\Gamma_{sub}^{max} $, which can be 
computed from Eq.~(\ref{gq}) given in Appendix~\ref{app:A}, setting $q=N-1$.

In Fig.~\ref{Ptth} we compared  numerical data for $P(t)$ with
the analytical expression given in Eq.~(\ref{Pth}): the agreement is
excellent for disorder strength $W < W_{cr}$, where the
 decay is well approximated by a bi-exponential function.
From Eq.~(\ref{Pth}) it is also possible
to compute the time at which a change in the decay occurs, $t^*$,
by equating  the two terms on the r.h.s.~of Eq.~(\ref{Pth}). 
Dividing by $2 \pi c$ in order to have $t^*$ in picoseconds, we obtain
\begin{equation}
t^* = \frac{1}{2 \pi c(N\gamma-\Gamma_{sub}^{max})  }
\log{\left(\frac{c_1}{1-c_1} \right )}\,.
\label{ts}
\end{equation}
Such a time,  for one value of the disorder strength, 
 is shown with an arrow in Fig.~\ref{Ptth}.
Note that $t^*$ can be considered as the time up to which the decay is
superradiant. As the disorder increases, $t^*$ goes to zero and the
decay of the extended state becomes similar to the decay of independent sites, {i.e.}~$P(t)=e^{-\gamma t}$.

\begin{figure}[t]
\vspace{0cm}
\includegraphics[width=8cm,angle=0]{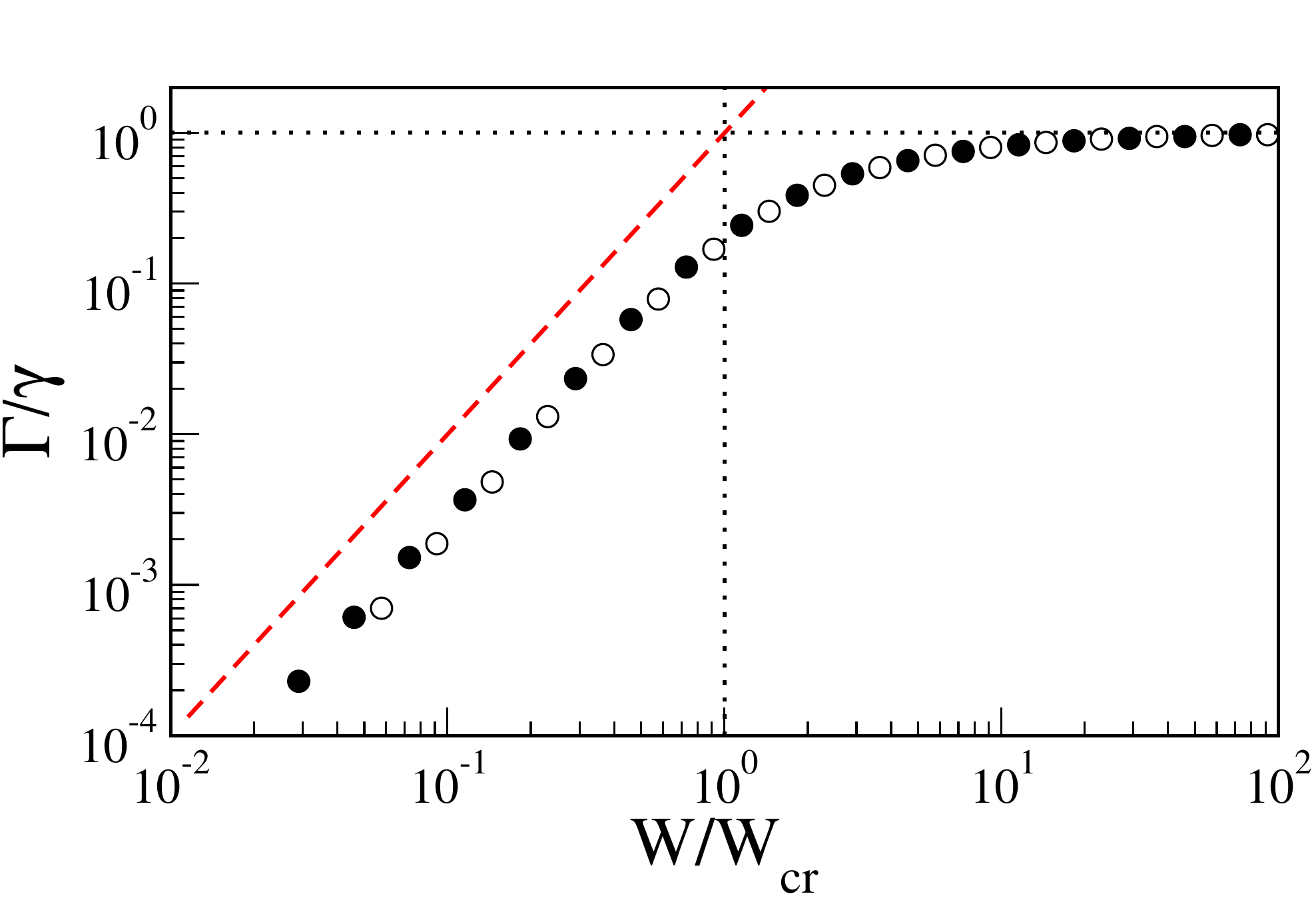}
\caption{(Color online)
Inverse Decay time of the average probability $\langle P(t) \rangle$, rescaled to
the individual decay width $\gamma$, as a function of the rescaled disorder strength
$W/W_{cr}$.
Different sets of values have been considered: full circles are for $N=20$, $\gamma=2.5$, $\Omega=0.5$,
$W_{cr}=86.6$; open circles are for $N=100$, $\gamma=0.1$, $\Omega=0.1$,
$W_{cr}=17.32$. The dashed red line represents $\langle \Gamma_{sub}\rangle =\gamma$.
}
\label{dyn}
\end{figure}

The generality of our results can be assessed by observing that 
the critical disorder at which Superradiance is quenched  is
an important threshold for the whole system dynamics and not only
for the superradiant state.
To this end, let us consider as initial state
 a random superposition of site states
$$
|\psi_0 \rangle = \sum_{k=1}^N c_k  | k \rangle,
$$
$c_k$ being random complex coefficients such that $\sum_{k=1}^N |c_k|^2 = 1$.
For such initial state, we compute 
the survival probability $P(t)$, for one realization of disorder.
By changing the random initial state and the random diagonal disorder,
we can consider the average survival probability $\langle P(t)\rangle$ and define
 the decay time $\tau\equiv 1/\Gamma$ as
$$
\langle P(\tau )\rangle = 1/e.
$$
These $\Gamma$ values are reported 
in Fig.~\ref{dyn}  as a function of $W/W_{cr}$
for different parameter values. For the sake of comparison
we show, in the same figure,
the analytical expression
for the average decay width of the subradiant states, see  Eq.~(\ref{Wcr}).
As one can see, up to a numerical constant, the agreement is very good.
In other words, the disorder strength necessary to quench Superradiance
(obtained analytically imposing the average decay width of the subradiant states, 
$\langle \Gamma_{sub} \rangle$,
to be equal to the single site decay $\gamma$)
is also a valid tool in estimating the decay time of the survival 
probability associated with generic random initial conditions.

The problem of computing the survival probability of the superradiant
state in presence of inhomogeneous broadening was also
 considered in~\cite{woggon} for $N$ two-level systems.
A bi-exponential behavior was numerically found for any 
 excitation number. For the case where only one
excitation is present, our results are compatible with the bi-exponential
behaviour of the survival probability found in~\cite{woggon}, and  we also  give
 an approximate analytical expression for the survival probability $P(t)$.

Finally, in order to stress the relevance of superradiant energy
transfer in natural photosynthetic complexes, let us consider the excitation transfer from
the LHI complex to the reaction center~\cite{schulten,schulten1}. 
First of all, we point out that all models used to study the dynamics
of this complex are characterized
by a large inhomogeneity in the transfer time of
different energy eigenstates~\cite{schulten,schulten1,fassioli}, which is, for instance,
typical of the  superradiant regime. Thus, we  reasonably  assume  Superradiance in
transfer to be  relevant in natural complexes. 
We can estimate $\gamma$, representing the coupling to the reaction center, from the following considerations.
Using realistic parameters
as was done at the end of Sec.~\ref{sec:3}, $\Omega \approx
600\,\mathrm{cm}^{-1}$, $N=32$, 
we computed  $P(t)$ for 
the fully extended  state of Eq.~(\ref{SR}), in presence 
of a realistic value of the static disorder $W=320\,\mathrm{cm}^{-1}$ (a large disorder corresponding to the critical disorder at which
Superradiance starts to be quenched).  
We choose $\gamma=0.01\,\mathrm{cm}^{-1}$, 
so that we have a transfer
time (time at which $P(t)=1/e$) starting from the fully extended
state of $\approx 35\,\mathrm{ps}$, in agreement with experimental
data~\cite{schulten1}. Note that a single occupied site would give a transfer
time of $500\,\mathrm{ps}$, showing that,  even in presence of strong and
realistic static disorder, Superradiance is able to strongly enhance
energy transfer.

\section{Conclusions}\label{sec:6}

We analyzed the interplay of Superradiance, induced by
the coupling to a common decay channel, and
Localization, induced by static disorder, in 1-d ring-like structures,
usually used to model some natural light-harvesting complexes. The
common decay channel can represent both the coupling to the
electromagnetic field or to a central core absorber, such as the
reaction center in natural photosynthetic complexes. 
We have shown that, for zero disorder, these 
structures are in a superradiant regime for any value of the coupling strength
to a common decay channel. 
Above a critical disorder strength 
superradiant effects decrease until,
for very large disorder, 
all of the states decay independently with the common width $\gamma$, 
and cooperativity is completely lost. 
Our main purpose was to determine the critical disorder at which
Superradiance is hindered.
Using a perturbative approach  we
determined analytically such  critical disorder  and we related it
firstly with the localization properties of superradiant and subradiant
states and then to the system dynamics.
We found that Superradiance can be quenched by disorder
in different ways, depending on the regime entailing either weak or strong coupling to the continuum. These regimes are triggered
by the parameter $N\gamma/4\Omega$, which represents the ratio
between the coupling strength to the continuum, $\gamma$, and the unperturbed
mean level spacing in absence of disorder, $4\Omega/N$. 
When this ratio
is small, {i.e.}~$N\gamma/4\Omega \ll 2(\pi/N)^2$,
 the critical disorder is independent of the coupling  
strength with the
external environment and it is determined only by the parameters of
the molecular chain, since the opening is unable to affect the disorder-induced Localization. 
In this regime, the critical disorder decreases
with the size of the system, but, for large system size $N\to \infty$,
such a regime becomes less and less feasible (to be in the 
weak opening regime implies the condition
$N^3 \ll 8\pi\Omega/\gamma  $).
On the other hand, for strong opening, $N\gamma/4\Omega \gg 1 $,
 the critical disorder increases with both the size of the system and the coupling  
strength with the external environment.  This is 
in agreement with  the results  recently found in Ref.~\cite{laltro}, 
where the same ring structure has been analyzed
in presence of dephasing (dynamical disorder) and the strength necessary
to destroy Superradiance was found to be 
proportional to both $\gamma$ and $N$.

We also demonstrated that the critical disorder at which Superradiance is 
suppressed is close to the disorder
at which superradiant states localize~\cite{alberto}.
Specifically, we found that, in the weak opening regime, $N\gamma/4\Omega \ll
2(\pi/N)^2$, Superradiance is quenched at the same disorder at which
the edge state of the closed system, with real energy equal to that of the 
superradiant state, localizes. 

We have also found that, in the strong opening regime, $N\gamma/4\Omega \gg 1 $, 
Superradiance is a manifestation of cooperative robustness to disorder, in that the 
superradiant state localizes at a disorder strength (proportional
to the system size) much larger than the one needed to localize the 
corresponding edge state of the closed system.
As for subradiant states, in any regime, 
they begin their process of localization
at the same disorder strength at which the states 
of the closed system do.

Finally, we have shown the relevance of our findings to natural
photosynthetic complexes: $i)$ for the realistic parameters of natural
complexes, Superradiance is quenched at a disorder strength which is
independent of the coupling to the external environment (electromagnetic
 field or reaction center), which is difficult to determine experimentally. Thus
our findings allow to determine the critical disorder from the
localization properties of the closed system alone; 
$ii)$ the critical
disorder thus obtained is compatible with experimental estimates, 
 suggesting that natural systems operate close to the Superradiance Transition; 
$iii)$ even in presence of large and realistic
static disorder, Superradiance can strongly enhance energy transfer to
the reaction center and light absorbtion.

{ \it Acknowledgments.}
We would like to thank R.~Kaiser and B.~Sterzi for providing many useful discussions.


\appendix
\section{Decay widths, a perturbative approach}
\label{app:A}

Perturbation theory is applied to the  \emph{symmetric
  unperturbed Hamiltonian,} 
$H^{tb} -i \gamma Q/2, $
 in order to  find the
 critical  
 disorder
strength, $W_{cr}$, at which Superradiance is destroyed. 
Let us rewrite the Hamiltonian given in Eq.~(\ref{amef})
as
$$
H_{\mathrm{eff}}= H^{tb}-\frac{i\gamma}{2} Q+D\,,
$$
where $ H^{tb}$ is the tight binding Hamiltonian, Eq.~(\ref{AM}), in absence of
 disorder and $D= \sum_i \epsilon_i |i\rangle \langle i|$, see Eq.~(\ref{dam}), is a diagonal matrix which 
contains the 
  disordered site energies $\epsilon_i $.

It is necessary to define the non-Hermitean ``bra'' as the transposed of a ket 
$$\bbra{\psi}:=(\ket{\psi})^t,$$ 
while the standard bra is the adjoint $$\bra{\psi}:=(\ket{\psi})^\dag.$$
 Indeed, given the right eigenvectors of a symmetric Hamiltonian $\ket{\psi_i}$, 
the left eigenvectors are $\bbra{\psi_i}$, that is
\[
H\ket{\psi_i}={\cal E}_i\ket{\psi_i}\,,\qquad\text{and}\qquad\bbra{\psi_i}H=
{\cal E}_i\bbra{\psi_i}\,,
\]
and we have the \emph{biorthogonality} condition
\[
\bbraket{\psi_i}{\psi_j}=\delta_{ij}\,.
\]
Clearly, for real eigenstates 
we have $\bbra{\psi}=\bra{\psi}$.

Matrix elements of the operators defined above, in the site basis
 $\{\ket{s},s=1,\ldots,N\}$,
are given by,
\begin{equation}
\label{eq:defH0}
D_{ss}=\epsilon_s\;,\quad H^{tb}_{1N}=H^{tb}_{ss+1}=-\Omega\;,\quad Q_{sr}=1\;,
\end{equation}
with $r,s=1,\ldots,N$. In Eq.~(\ref{eq:defH0}) 
$\Omega>0$, and
 $-W < \epsilon_s < W $, are independent identically distributed random 
variables with mean $0$ and variance $W^2/12$. 

Since
$[H^{tb},Q]=0,$
it is convenient to study the whole system on the basis of eigenstates of $H^{tb}$, which are given by
\[
\braket{s}{\psi_q}=\frac{1}{\sqrt{N}} \cos { \frac{2 \pi sq}{N}}
\]
for $q=1,\ldots,N/2,N$, and 
\[
\braket{s}{\psi_q}=\frac{1}{\sqrt{N}} \sin { \frac{2 \pi s(N-q)}{N}}
\]
for $q=N/2+1,\ldots,N-1$,
with eigenvalues
\[
e_q=-2\Omega\cos\frac{2\pi q}{N}\,.
\]
The eigenvalues of the Hamiltonian, $H^{tb}-(i/2)\gamma Q$,
 are thus given by,
\begin{equation}
\label{ceig}
\eps_q=-2\Omega\cos\frac{2\pi q}{N}-(iN\gamma/2) \delta_{qN},
\end{equation}
that is, only the ground state acquires a decay width $N\gamma$. Such a state is called \emph{superradiant}, and the others are \emph{subradiant}. Notice that $\ket{\psi_N}$ 
and $\ket{\psi_{N/2}}$ are non-degenerate, 
while, for any $q=1,\ldots,N/2-1$, $\ket{\psi_q}$ and $\ket{\psi_{N-q}}$ 
span a two-dimensional degenerate eigenspace. 

When the disorder strength is turned on
 every state will  get an eigenenergy with a negative
 imaginary part (decay width).
Perturbation theory up to second order can be applied, 
for  sufficiently small disorder strength, to give
\begin{align}
&\mbox{}\eps'_q=\eps_q+\bbra{\psi_q}D\ket{\psi_q}+\sum_{q'\neq q}\frac{\bbra{\psi_q}D\ket{\psi_{q'}}^2}
{\eps_q-\eps_{q'}}\notag\\
&\mbox{}=\eps_q+\sum_{s=1}^N\epsilon_s\braket{s}{\psi_q}^2\label{eq:gamma}\\
&\mbox{}+\sum_{s,s'=1}^N\sum_{q'\neq q}
\frac{\epsilon_s \epsilon_{s'}\bbraket{\psi_q}{s}\braket{s}{\psi_{q'}}\bbraket{\psi_{q'}}{s'}
\braket{s'}{\psi_q}}{\eps_q-\eps_{q'}}\,.\notag
\end{align}
For degenerate energy levels the first order
 correction is given by the eigenvalues of the 2 by 2 symmetric matrices
\begin{multline}
\begin{pmatrix}
\bbra{\psi_q}D\ket{\psi_q} & \bbra{\psi_q}D\ket{\psi_{N-q}} \\
\bbra{\psi_{N-q}}D\ket{\psi_q} & \bbra{\psi_{N-q}}D\ket{\psi_{N-q}} 
\end{pmatrix}=\\
=\begin{pmatrix}
\sum_{s}\epsilon_s\braket{s}{\psi_q}^2 & \sum_{s}\epsilon_s\braket{s}{\psi_{q'}}\braket{s}{\psi_q} \\
\sum_{s}\epsilon_s\braket{s}{\psi_{q'}}\braket{s}{\psi_q} & \sum_{s}\epsilon_s\braket{s}{\psi_{q'}}^2 
\end{pmatrix}
\end{multline}
while the second order correction is
\[
\sum_{s,s'=1}^N\sum_{q'\neq q,N-q}\frac{\epsilon_s\epsilon_{s'}\bbraket{\psi_q}{s}
\braket{s}{\psi_{q'}}\bbraket{\psi_{q'}}{s'}\braket{s'}{\psi_q}}{\eps_q-\eps_{q'}}\,.
\]
We are interested in the imaginary part of the perturbed eigenvalues, and, since the 
eigenstates, $\ket{\psi_q}$, are \emph{real}, first order corrections never contribute to those terms.
Considering now averages over disorder
and writing $\eav{\eps'_q}=\delta_q-i\gamma_q/2$, 
with $\delta_q,\gamma_q$ real, we obtain the average decay widths
for the superradiant state
\begin{align}
\gamma_N&\mbox{}=N\gamma-\gamma\frac{W^2}{48\Omega^2N}\sum_{s=1}^N \sum_{q'=1}^{N-1}\frac{1}{\left(1-\cos\frac{2\pi q'}{N}\right)^2+\frac{N^2\gamma^2}{16\Omega^2}}\notag\\
&\mbox{}= N\gamma-\frac{\gamma W^2}{48\Omega^2}\sum_{q'=1}^{N-1}\frac{1}{\left(1-\cos\frac{2\pi q'}{N}\right)^2+\frac{N^2\gamma^2}{16\Omega^2}}\,,
\end{align}
and, for the subradiant ones, $q=1,\ldots,N-1$,
\begin{align}
\gamma_q&\mbox{}=\frac{\gamma
  W^2}{48\Omega^2N}\sum_{s=1}^N\frac{1}{\left(\cos\frac{2\pi
      q}{N}-1\right)^2+\frac{N^2\gamma^2}{16\Omega^2}}\notag\\
&\mbox{}= \frac{\gamma
  W^2}{48\Omega^2}\frac{1}{\left(\cos\frac{2\pi
      q}{N}-1\right)^2+\frac{N^2\gamma^2}{16\Omega^2}}\,.
\label{gq}
\end{align}
The maximum decay widths of the subradiant states are clearly $\gamma_{1}$ and $\gamma_{N-1}$, while the average of the subradiant widths is
\begin{equation}
\label{pte}
\langle \Gamma_{sub} \rangle=\frac{\gamma W^2}{48\Omega^2(N-1)}\sum_{q=1}^{N-1}\frac{1}{\left(\cos\frac{2\pi q}{N}-1\right)^2+\frac{N^2\gamma^2}{16\Omega^2}}\,.
\end{equation}
We finally define the critical disorder $W_{cr}$ as the one at which
 $$\langle\Gamma_{sub}\rangle = \gamma, $$ 
{i.e.}~equals the single-site  decay width $\gamma$.

Let us now apply  first order perturbation theory to the superradiant
state. For $W=0$, the superradiant state is given by the extended
state, Eq.~(\ref{SR}), while,  for $W \ne 0$, we 
can write
$$
|SR \rangle \simeq (1/\sqrt{N}) \sum_{k=1}^N |k\rangle + \sum_{q \ne 1} \frac{D_{1,q}}{\epsilon_1-\epsilon_q} |\psi_q \rangle\,.
$$
From this expression we can compute the probability to be in the
superradiant state when starting from the extended state as
\begin{equation}
c_1=\dfrac{1}{1+\dfrac{W^2}{48 \Omega^2 N}\displaystyle{ \sum_{s=1}^{N-1}}
  \dfrac{1}{(1+\cos(2 \pi s/N))^2+(\gamma N/4 \Omega)^2} }\,.
\label{c1}
\end{equation}

\section{Approximate formula for perturbative average width}\label{app:B}

Let us rewrite  Eq.~(\ref{pte})  in the form
\begin{equation}
\langle{\Gamma}_{sub}\rangle=\frac{\gamma W^2}{48\Omega^2 }  S_a\,,
\label{ptes}
\end{equation}
where we have defined
\begin{equation}
S_a  = \frac{1}{N-1} \sum_{q=1}^{N-1}\frac{1}{\left(\cos\frac{2\pi q}{N}-1\right)^2+a^2}\,,
\label{essea}
\end{equation}
and
$$
a=  \dfrac{N\gamma}{4\Omega}\,.
$$

Eq.~(\ref{essea}) can be put in integral form for sufficiently large $N$, that is $2\pi/N \ll 1 $, as
\begin{equation}
\begin{aligned}
S_a&\mbox{}  = \frac{1}{2\pi } \int_0^{2\pi} \ dx \frac{1}{\left(\cos x -1\right)^2+a^2}\\
&\mbox{} =\frac{\left( 2a + 2\sqrt{4+a^2} \right)^{1/2}}{a^{3/2} \sqrt{4+a^2}}\,.
\end{aligned}
\label{esseaint}
\end{equation}
It is easy to show that Eq.~(\ref{esseaint}) has two different limits, namely 
\begin{equation}
S_a = \left\{
\begin{array}{cl}
1/(2 a^{3/2} )  &   \quad  {\rm for} \;   a \ll 1 \,, \\
& \\
     1/a^2  &  \quad  {\rm for}  \; a \gg 1 \, .
     \end{array}
\right.
\label{asymp}
\end{equation}

Nevertheless, substituting  a sum with an integral works only for very large $N$. For small $N < 100$,
or, in general for a sufficiently small $a$ value, it is more convenient approximating the sum 
with only two terms, namely those for which the denominator in Eq.~(\ref{essea}) is small. In detail, 
one has 
\begin{equation}
1- \cos (2\pi/N) \approx 2\pi^2/N^2 \,
{\rm for}\, a < a_{cr} = 2\pi^2/N^2.
\end{equation}

This implies that, in this regime,
$$
S_a = const \approx N^3/2\pi^4\,.
$$
On the other hand, for $ a_{cr} < a < 1 $ (for all those $N$ values for which  $ a_{cr} < 1 $),
one can approximate the sum with the integral and use the asymptotic behavior 
given in Eq.~(\ref{asymp}). 
To summarize we have the following behavior:
\begin{equation}
S_a = \left\{
\begin{array}{cl}
\dfrac{N^3}{2 \pi^4}   &  \quad  {\rm for}  \; a \ll  \dfrac{2\pi^2}{N^2} \,, \\
      & \\
\dfrac{1}{2 a^{3/2}}   &  \quad  {\rm for}  \;  \dfrac{2\pi^2}{N^2} < a <  1 \,, \\
      & \\
\dfrac{1}{a^2}   &   \quad  {\rm for} \;   a \gg 1\,. 
     \end{array}
\right.
\label{asymp1}
\end{equation}

These different regimes can be written in terms of physical parameters as follows:
\begin{equation}
\langle {\Gamma}_{sub}\rangle  = \left\{
\begin{array}{cl}
\displaystyle \frac{ \gamma W^2 N^3}{ 96 \pi^4 \Omega^2} & \quad {\rm for} \; 
\displaystyle   \frac{N^3 \gamma}{8\pi^2 \Omega} \ll 1  \,,\\
      & \\
\displaystyle \frac{ \gamma W^2 }{ 12\gamma^{1/2}  \Omega^{1/2}N^{3/2}   } & \quad {\rm for} \; 
\displaystyle  \frac{2\pi^2}{N^2} \ll \frac{N \gamma }{4 \Omega }  \ll 1 \,, \\  
      & \\
\displaystyle \frac{ W^2 }{ 3 N^2 \gamma    } & \quad {\rm for} \;
\displaystyle   \frac{N \gamma }{4 \Omega }  \gg  1 \,.      
     \end{array}
\right.
\label{asymp2}
\end{equation}

\end{document}